# Title

**Computational comparison of the bending behavior of aortic stent-grafts**

# Authors

**Nicolas Demanget[1]; Stéphane Avril[1]; Pierre Badel[1] ; Laurent Orgéas[2]; Christian Geindreau[2]; Jean-Noël Albertini[3] ; Jean-Pierre Favre[3]**

[1]Ecole Nationale Supérieure des Mines de Saint-Etienne, Center for Health Engineering, CNRS UMR 5146, INSERM SFR IFRESIS, Saint-Etienne F-42023, France. [2]Grenoble-INP, UJF-Grenoble 1, CNRS UMR 5521, 3SR Lab, Grenoble F-38041, France. [3]CHU Hôpital Nord, Department of Vascular Surgery, Saint-Etienne F-42055, France.

# Corresponding author

**Nicolas Demanget**

158 cours Fauriel F-42023 Saint-Etienne cedex 2

Tel: +33(0)477499772

E-mail address: nicolas.demanget@gmail.com

# ABSTRACT

Secondary interventions after endovascular repair of abdominal aortic aneurysms are frequent because stent-graft (SG) related complications may occur (mainly endoleak and SG thrombosis). Complications have been related to insufficient SG flexibility, especially when devices are deployed in tortuous arteries. Little is known on the relationship between SG design and flexibility. Therefore, the aim of this study was to simulate numerically the bending of two manufactured SGs (Aorfix - *Lombard Medical* (A) and Zenith – *Cook Medical Europe* (Z)) using finite element analysis (FEA). Global SG behavior was studied by assessing stent spacing variation and cross-section deformation. Four criteria were defined to compare flexibility of SGs: maximal luminal reduction rate, torque required for bending, maximal membrane strains in graft and maximal Von Mises stress in stents. For angulation greater than 60°, values of these four criteria were lower with A-SG, compared to Z-SG. In conclusion, A-SG was more flexible than Z-SG according to FEA. A-SG may decrease the incidence of complications in the setting of tortuous aorto-iliac aneurysms. Our numerical model could be used to assess flexibility of further manufactured as well as newly design SGs.

# TEXT

## 1. Introduction

Abdominal aortic aneurysms represent a major public health issue. Aneurysm rupture causes 15000 deaths per year in the USA. Prevention of AAA rupture is based on surgical treatment. The traditional approach consists in opening the patient's abdomen and replacing the aneurysmal portion of aorta by sewing a prosthetic graft. However, this technique is invasive and carries relatively high mortality and morbidity, especially in high risk patients.

A less invasive alternative technique, endovascular aneurysm repair (EVAR), has been developed over the last twenty years. This technique consists in excluding the aneurysm sac from the main stream circulation by the endovascular insertion of a stent-graft (SG) via the femoral arteries. Most current SGs use a combination of metallic stents stitched to a polymeric fabric. EVAR reduces postoperative morbidity and mortality when compared to open repair (Greenhalgh et al., 2004). However, SG durability remains the principal issue. The incidence of complications requiring secondary interventions, such as endoleaks (Albertini et al., 2001; Baum et al., 2003) and stenosis or thrombosis of the SG (Caroccio et al., 2002; Cochennec et al., 2007), increases during follow-up (Greenhalgh et al., 2010). Clinical studies suggest that insufficient flexibility of SG may induce kinks within tortuous iliac arteries (Umsheid et al, 1999, Caroccio et al., 2002). The lack of flexibility in these studies has been related to the rigid Z-stents supporting the SG (Figure 1A). To overcome this problem, a new generation of more flexible SGs has been developed (Hinchliffe 2004, Saratzis 2008). The increased flexibility was obtained by the use of ring or spiral shaped nitinol stents (Figure 1B). Although no study focused specifically on the assessment of SG flexibility, one non randomized clinical study has suggested that the use of a newly designed more flexible SG decreased the incidence of limb thrombosis (Weale et al., 2010). However, no study has provided yet a quantitative analysis of SG flexibility. Durability may also be compromised by fabric tear (Chakfe et al., 2004) and stent fracture (Zarins et al., 2004). Many studies have been conducted on separate assessment of fabric (Heim et al., 2009) and stents (Pelton et al., 2008) durability. However, we found no study combining stent and fabric durability assessment. Kleinstreuer et al. (2008) performed a finite element analysis to assess various aspects of SG mechanics, but they used a general design for stents and fabric that did not reproduce any of the current manufactured devices. Furthermore, no evaluation was performed in a tortuous setting in this study.

The aim of this study was to design numerical models of Z-stented and spiral stented SG limbs, in order to provide a quantitative comparison of their bending behavior.

## 2. Methods

Two manufactured SG limbs were modeled using a finite element mechanical analysis software (SIMULIA Abaqus 6.8/Explicit®): Aorfix (A-SG) *(Lombard Medical, Didcot, United Kingdom)* and Zenith (Z-SG) *(Cook Medical Europe, Bjaeverskov, Denmark)* (Figure 1).

### 2.1. Stent-grafts

SG iliac limbs were chosen because they are usually subjected to important deformations within the iliac arteries. Limb samples were obtained from the manufacturers.

#### 2.1.1 Zenith

This first generation SG has been developed since 1990 and subsequently implanted in human in 1993. Iliac limbs are composed of 316L stainless steel Z-stents attached with polypropylene running sutures to a polyester woven fabric (Dacron). Dimensions of the sample were as follows: proximal and distal diameter 12 and 16 mm, respectively; length 111 mm. Proximal and distal stents were internal to the fabric and their length was 21 and 17 mm, respectively. Four intermediate stents were external to the fabric and their length was 12 mm.

#### 2.1.2 Aorfix

This new generation SG has been developed since 1998 and first implanted in human in 2001. Iliac limbs are composed of a continuous external spiral nitinol (NiTi) stent attached with polyester running sutures to a polyester woven fabric (Dacron). Dimensions of the sample were as follows: proximal and distal diameter 12 and 16 mm, respectively; length 110 mm.

## 2.2. Stent-graft modeling and simulation of bending

### 2.2.1 Geometry and mesh

In order to make consistent comparison between A-SG and Z-SG, dimensions and design of the modeled stent-grafts were slightly modified from the original samples (Figure 1). Dimensions of the modeled grafts (*i.e.* prosthetic fabric) and stents are reported in Table 1. While cylindrical grafts were modeled and meshed with Abaqus 6.8.2, we have used Matlab R2009a to generate geometries and meshes of stents. It was necessary to have correct contact geometries for the numerical simulation.

#### *2.2.1.1 Stents*

By giving the number of discretization nodes *nnod*, the number of patterns *npat*, the stent height *sth* and the stent radius *str*, the Matlab routine generated the mesh nodes of a Z-SG stent by using a cylindrical system (Figure 2). The angle $\beta$, which was defined as the angle between the first and the current node in *xy*-plane, was incremented from 0 to $2\pi$. Each triplet of coordinates *x*, *y* and *z* was generated according to the following equations:

$$x = str \times \cos \beta \quad (1)$$

$$y = str \times \sin \beta \quad (2)$$

$$z = sth/2 \times \sin(\beta \times npat) \tag{3}$$

For A-SG, the routine was modified in order to generate a helical stent. Thus, *nnod*, *str*, the number of turns *ntu* and a spacing factor between rings *spf* were defined. Finally, node coordinates were generated using the equations (1) and (2) as well as the following one:

$$z = spf \times \beta \tag{4}$$

Mesh connectivity was also generated with the above mentioned program.

### 2.2.1.2 Fabric and sutures

Graft geometries were meshed with linear shell elements (S3 in Abaqus) while beam elements (B31) were chosen for the stents. The number of elements and degrees of freedom (DOF) are presented in Table 1. Because stent and graft are sutured together, bonding between the graft outer surface and the outer surface of the metal scaffolding was imposed ("tie constraint" in Abaqus) so that stents and graft could not slide or separate during simulations. Sutures which secure metal scaffolding on graft were not considered in the models. A self-contact algorithm was added in order to avoid self-penetration of components.

## 2.2.2 Material properties

### 2.2.2.1 Stents

An elastoplastic isotropic model with parameters taken from literature (Aurrichio et al., 2001) and listed in Table 2 was used to model the mechanical behavior of 316L stainless steel Z-SG stents.

The material model for Nitinol was also considered as elastoplastic and isotropic. Thus, the hardening of stress-strain curves reproducing the forward stress induced martensitic transformation observed in NiTi alloys. It was assumed that loading of any material point in the stent was monotonic during the simulation of SG bending. This strong assumption had been checked *a posteriori*: no unloading was observed within the Nitinol stents. Furthermore, the tension/compression asymmetry, often observed in NiTi alloys (Orgéas and Favier, 1995, 1998), was neglected in this study. Parameters of this NiTi beam model are listed in Table 3. These material properties were taken from Kleinstreuer et al (2008).

*2.2.2.2 Fabric*

Data available in the literature regarding mechanical properties of the polyester fabric of Z-SG and A-SG were considered insufficient. Therefore, polyester samples were obtained from a manufacturer of SG fabric (Vascutek, Inchinnan, United Kingdom).

A preliminary study (results not shown here) demonstrated the need to consider both the anisotropy and the bending stiffness of the fabric. Thus, two inputs were necessary to implement the fabric material model: the in-plane elastic behavior and the bending behavior.

- In-plane elastic behavior

The in-plane elastic behavior of the fabric was assessed by performing pure and plane strain tensile tests using a tensile testing machine (Gabo Eplexor 500, Ahlden, Germany). The machine was set on static mode, with a load cell of 25 N. Small rectangular fabric samples (3 mm

x 30 mm) were cut. Different orientations $\theta$ of samples with respect to tensile direction were tested, from 0° (longitudinal or warp direction) to 90° (circumferential or weft direction), with increments of 11.25° (Figure 3). Four samples were tested for each orientation. In order to determine the value of the Poisson's ratio, plane tensile tests were performed on other samples. Each sample was tested until rupture, using a crosshead speed of 0.1 mm/min.

The strong anisotropy observed on tensile stress-strain curves required considering an elastic orthotropic plane stress model for the in-plane behavior. This model was defined by the following constitutive equation expressed in the local coordinate system $(\vec{e_L}, \vec{e_C})$:

$$\begin{pmatrix} \varepsilon_{LL} \\ \varepsilon_{CC} \\ 2\varepsilon_{LC} \end{pmatrix} = \begin{pmatrix} 1/E_L & -\dfrac{\upsilon_{LC}}{E_L} & 0 \\ -\dfrac{\upsilon_{CL}}{E_C} & 1/E_C & 0 \\ 0 & 0 & 1/G_{LC} \end{pmatrix} \begin{pmatrix} \sigma_{LL} \\ \sigma_{CC} \\ \sigma_{LC} \end{pmatrix} \qquad (5)$$

with $E_L$ the longitudinal elastic modulus, $E_C$ the transversal elastic modulus, $\nu_{LC}$ and $\nu_{CL}$ the major and minor Poisson's ratio and $G_{LC}$ the shear modulus.

The fabric behavior was assumed to be linear in each direction and elastic moduli were assessed for each angle by calculating the secant modulus for a tensile strain of 10%, *i.e.* within the nearly linear portion of the stress-strain curves. The averaged values of $E_\theta$ obtained from tensile tests are plotted on the graph of Figure 4 as a function of $\theta$. This graph clearly showed the strong in-plane orthotropy of the textile. Error bars also shown in the graph corresponded to the maximal scattering observed with the tested samples. Then, a Matlab routine was used to assess the value of the shear modulus $G_{LC}$ from the knowledge of $E_\theta$ and $\nu_{LC}$ and by using the following equation:

$$\frac{1}{E_\theta} = \frac{\cos^4\theta}{E_L} + \frac{\sin^4\theta}{E_C} + \frac{1}{4}\left(\frac{1}{G_{LC}} - \frac{2\upsilon_{LC}}{E_L}\right)\sin^2(2\theta) \qquad (6)$$

The resulting fit shown in Figure 4 was satisfactory. Material properties obtained for the in-plane behavior of tested fabric are listed in Table 4.

- Bending behavior

The bending behavior of the fabric was evaluated by an inverse method that combined numerical simulation and the "nail test". The latter consisted in clamping one edge of the fabric sample (slender ribbon) and leaving the other tip free, so that the sample could bend under its own weight. In addition, this test was simulated numerically by modeling the fabric with shell elements.

The aim of this inverse method was to find the most appropriate material parameters so as to reproduce the experimental curvature of the sample with the simulation while taking care not to change the in-plane behavior of the fabric. Accordingly, elastic moduli together with the equivalent thickness $t$ of the fabric were artificially modified in order to adjust the bending rigidity of the shell elements.

Indeed, according to the Kirchhoff-Love theory, the bending rigidity D of a shell element is defined by the following equation:

$$D = \frac{E_L t^3}{12(1-\upsilon^2)} \qquad (7)$$

with $E_L$ the elastic modulus, $t$ the thickness of the shell element and $\upsilon$ the Poisson's ratio. In our case, the graft model was orthotropic. Thus, there were two bending rigidities, one in the longitudinal direction $D_L = 4.10^{-4}$ N.mm and the other in the circumferential direction $D_C = 18.10^{-4}$ N.mm. Modeling the fabric with shell elements without artificially adjusting the value of thickness and elastic moduli would have led to $D_L = 1.10^{-2}$ N.mm and $D_C = 4.10^{-2}$ N.mm, which was far from the actual values of this fabric.

The orthotropic model was finally implemented in the Abaqus software by using a "Lamina" material model, *i.e.* an orthotropic plane stress model.

### 2.2.3. Simulation of bending (boundary conditions)

The bending response of each SG was computed using boundary conditions sketched in Figure 5. The stent or the portion of stent at each graft extremity was considered a rigid body controlled by a reference point. Motions were applied directly onto these two rigid bodies. Opposite rotations were applied about the *x*-axis, until a bending angle, $\alpha$, of 180° was reached. The maximum value of 180° for $\alpha$, which determined the magnitude of SG deformations, was chosen on the basis of the following clinical data. Iliac artery angulations have been studied in a number of publications reporting the clinical results of AAA treatment using flexible SGs (Weale et al., 2010; Balasubramaniam et al., 2009; Perdikides et al., 2009). The proportion of patients with iliac angulation greater than 90° was up to 38%, with a maximum of 120° (Balasubramaniam et al. 2009). Our surgical team had experience of iliac angulation up to 180° (Albertini et al., 2006) (Figure 6).

In order to maintain the SG in the *yz*-plane, the other two rotations were locked. To avoid rigid body motions, two translations were also locked along the *x* and *y* axes. Translation along the *z*-axis (the initial longitudinal axis of the SG) was left free so as to avoid spurious tension in the longitudinal direction.

An explicit scheme was preferred to avoid convergence issues, since the analysis involved complex geometric, material and, especially, contact nonlinearities. In order to remain in a quasi-static case, dynamic effects were kept to a maximum of 5% of the static effects (Kim et al., 2002).

**2.3. Judgment criteria**

SG global behavior during bending was assessed using calculation of stent spacing, as well as shape change of SG cross-section in the $x'y'$-plane.

SG flexibility was assessed according to two criteria: maximal luminal reduction rate ($LR_{max}$) and torque required for bending the device ($TRB$). Mechanical behavior of stents and graft at the local scale was defined using two criteria: maximal Von Mises stress in the stents ($\sigma_S^{max}$) and maximal logarithmic membrane strain in the graft for longitudinal ($\varepsilon_{LG}$) and circumferential ($\varepsilon_{CG}$) directions. These four criteria were expressed as a function of the bending angle $\alpha$. Results were post-processed with Matlab R2009a.

### 2.3.1 Stent spacing variation

A-SG nodes were picked on inner and outer curvature of two consecutive stent turns in the middle of the SG (points A, B, A' and B') (Figure 7). Z-SG nodes on outer and inner curvature were picked up at the apex of distal strut of stent #2 (points C and C') and corresponding proximal strut of stent #3 (points D and D'). Distance between each pair of nodes was then calculated for $\alpha = 0°$, $\alpha = 90°$ and $\alpha = 180°$.

### 2.3.2 Cross-section shape change in *x'y'*-plane

The cross-section shape change was estimated in the middle of each device (Figure 7). Major and minor axis of corresponding polygon ($d_{x'}$ and $d_{y'}$, respectively) were then measured in the $x'y'$-plane for $\alpha = 0°$, $\alpha = 90°$ and $\alpha = 180°$. The shape change criterion $d_{x'}/d_{y'}$ was then defined to characterize the approximate distortion of the SG cross-sections.

### 2.3.3 Luminal reduction rate (*LR*)

*LR* of SG cross-section was defined as the reduction of SG cross-sectional area between initial ($S_0$ corresponding to $\alpha = 0$) and deformed state ($S$ corresponding to $\alpha > 0$):

$$LR = 100 \times (1 - \frac{S}{S_0}) \quad (\%) \tag{7}$$

This criterion characterizes the change of SG cross-sectional area which was computed as follows. Cross-section plane perpendicular to the longitudinal axis of the SG was defined at a given point along the SG and for a given value of $\alpha$ (Figure 8A). SG nodes were tracked down within this plane (Figure 8B). The area of the polygon circumscribed by the tracked nodes was then calculated.

*LR* was computed for 100 cross-sections along the longitudinal axis of the SG, and for 20 increasing values of $\alpha$ (Figure 8C).

Maximum *LR* ($LR_{max}$) was defined as the highest value obtained among the 100 cross-sections observed for a given value of $\alpha$. $LR_{max}$ was then plotted as function of $\alpha$ for each SG. Finally, *LR* was plotted for $\alpha = 180°$ as a function of the location of the cross-section along each SG centerline.

### 2.3.4 Torque required for bending (*TRB*)

*TRB* represents the torque required to bend the SG. *TRB* was obtained from the reaction moments picked up at the reference points at each extremity of the SG.

### 2.3.5 Stresses in stent ($\sigma_S^{max}$)

Stresses in stent were assessed using maximal Von Mises stresses ($\sigma_S^{max}$) directly generated by Abaqus. This criterion takes into account tension/compression, bending as well as torsion of the stents.

### 2.3.6 Strains in fabric ($\varepsilon_{LG}$ and $\varepsilon_{CG}$)

Longitudinal membrane strain ($\varepsilon_{LG}$) and circumferential membrane strain ($\varepsilon_{CG}$) were calculated by averaging out values of membrane strains generated by Abaqus for inner and outer surfaces of the shell elements. For that purpose, a local coordinate system ($\vec{e_L}, \vec{e_C}$) was defined along the yarn directions in order to ensure that output values corresponded to $\varepsilon_{LG}$ and $\varepsilon_{CG}$.

# 3. Results

## 3.1. Assessment of device global deformation

Figure 7 shows both devices for $\alpha=0°$, $\alpha=90°$ and $\alpha=180°$. For each value of $\alpha$, points A, B, A' and B' (C, D, C' and D', respectively) were plotted and used to assess stent spacing variation for A-SG (Z-SG, respectively) on inner and outer curvatures. On each representation of SG cross-section, distances $d_{x'}$ and $d_{y'}$ were measured. They allowed to assess distortion of the cross-section in *x'y'*-plane.

A-SG global deformation was homogeneous for any value of $\alpha$. Distance between stent turns on the inner curvature slightly decreased, whereas distance between turns remained constant on the outer curvature.

Z-SG global deformation may be described in two phases:

- Phase 1: From 0° to 90°, there was very little stent deformation and distance between stent struts on the inner curvature decreased gradually and significantly, whereas distance between struts remained constant on the outer curvature.

- Phase 2: From 90° to 180°, there was overlap of stent struts on the inner curvature and progressive collapse along the *y'*-axis of stent #3, as well as the adjacent parts of stent #2 and #4.

Figure 9 shows consistency of global deformation at 180° between actual samples and numerical models.

### 3.1.1 Stent spacing variation

Values of AB, A'B', CD and C'D' at corresponding values of $\alpha$ are reported in Table 5. A'B' and C'D' remained approximately constant for $\alpha$ between 0° and 180° since the stresses generated by SG bending were not sufficient to stretch the fabric in the outer curvature. On the contrary, AB and CD distances decreased with increasing values of $\alpha$. For A-SG, AB decreased from 4.4 mm to 1.7 mm because stent turns got closer in the inner curvature. For Z-SG, when $\alpha$ increased, stents #2 and #3 got closer and then overlapped, leading to a high reduction of CD.

### 3.1.2 Cross-section shape change in *x'y'*-plane

Values of $d_{x'}$, $d_{y'}$ and $d_{x'}/d_{y'}$ for both SGs at corresponding values of $\alpha$ are reported in Table 6. For A-SG, no important distortion was observed along the device for any value of $\alpha$. The cross-section for $\alpha = 180°$ flattened out slightly along the *y'*-axis. For Z-SG, $d_{x'}/d_{y'}$ barely increased from 1.0 to 3.1 for increasing values of $\alpha$, ie Z-SG cross-section progressively collapsed along the *y'*-axis and became almost oval for $\alpha = 180°$.

### 3.2. Luminal reduction rate (*LR*)

Figure 10A shows $LR_{max}$ plotted for each device as a function of $\alpha$. For both SGs, $LR_{max}$ increased with $\alpha$. $LR_{max}$ was similar in both SGs below 30°. For angles greater than 30°, $LR_{max}$ was greater for Z-SG, compared to A-SG. With A-SG, $LR_{max}$ stopped to increase above 60°, reaching a peak value of 14.6%, whereas with Z-SG, $LR_{max}$ continued to increase up to 180°, reaching a peak value of 80%.

Figure 10B represents the evolution of *LR* as a function of the location of the cross-section along the SG centerline, with an angle of 180°. For A-SG, *LR* values oscillated between 5% and 14.6% with a mean of 9.3%. Oscillations reflected graft wrinkling between spirals. For Z-SG, four peaks were visible along the SG. Each peak corresponded to a space between two stents. Larger deformations occurred in spaces between stents #2 and #3 and stents #3 and #4, with an asymmetrical distribution (respectively 80% and 60%).

### 3.3. Torque required for bending (*TRB*)

*TRB* for A-SG and Z-SG as a function of $\alpha$ is depicted in Figure 10C. For both SGs, *TRB* reached a value of around 4 N.mm from the beginning of bending. For A-SG, *TRB* value stayed relatively stable for $\alpha$ up to 180°. For Z-SG, *TRB* value stayed stable for $\alpha$ up to 80° and then increased in roughly linear fashion up to 180° to reach values up to 14 N.mm.

### 3.4. Stresses in stents ($\sigma_S^{max}$)

$\sigma_S^{max}$ as a function of $\alpha$ is presented in Figure 10D. For A-SG, $\sigma_S^{max}$ increased to 100 MPa between 0 and 30°, and stayed constant up to 150°; then $\sigma_S^{max}$ increased again from 100 to 177 MPa between 150° and 180°. For Z-SG, $\sigma_S^{max}$ increased linearly to 200 MPa at 100° and remained constant up to 180°. Therefore, $\sigma_S^{max}$ was equivalent for both SGs for low angulations (0° < $\alpha$ < 45°), and about twice as much for Z-SG for high angulations (90° < $\alpha$ < 150°). It must be pointed out that during A-SG bending, the NiTi stent remained in its elastic domain, since $\sigma_S^{max}$ was twice as small as the stress required to induce the forward martensitic transformation. Besides, Figure 10D also proves that plasticity may occur in the 316L stainless steel stents of Z-SG.

### 3.5. Strains in fabric ($\varepsilon_{LG}^{max}$ and $\varepsilon_{CG}^{max}$)

$\varepsilon_{LG}^{max}$ and $\varepsilon_{CG}^{max}$ as a function of $\alpha$ are presented in Figure 10E. For A-SG, both values remained inferior to 3% for any value of $\alpha$. For Z-SG, both values remained inferior to 3% for $\alpha < 120°$. For $\alpha > 120°$, values increased linearly, reaching a maximum of 11% for $\varepsilon_{LG}^{max}$ and 6% for $\varepsilon_{CG}^{max}$. FEA model of Z-SG showed that areas of maximum strains were located at the inner curvature, where stent struts overlapped each other (Figure 11). Finally, ultimate strain (see Table 4) was never reached for both SG fabrics.

## 4. Discussion

The results confirmed that stent design strongly influenced SG deformation during bending. Spiral stent of A-SG allowed homogeneous, low level deformation along the entire length of the SG. There was little movement between spirals as attested by low variations of AB and A'B'. On the contrary, Z-stents were responsible for heterogeneous two-phased response. Strong steel wires of Z-stents only allowed radial deformation, longitudinal deformation requiring much higher forces than the ones at stake in clinical setting. These properties accounted for the two-phased deformation of Z-SG during bending. Phase 1 was characterized by reduction of stent spacing on the inner curvature, while stent spacing on the outer curvature remained constant, together with minimal stent deformation. Beyond 90°, stent struts started to overlap and fabric resistance prevented further graft deformation at this point. Therefore, further load resulted in radial deformation of stents which was characterized by reduction of stents diameter predominantly along the $y'$-axis. This reduction of diameter accounted for the much higher $LR_{max}$ values recorded with Z-SG compared to A-SG.

$LR_{max}$ greater than 70% may favor SG thrombosis. It is well accepted in clinical practice that cross-section reduction greater than 70% of an artery increases the risk of thrombosis. These data are consistent with the results of Weale et al. (2010) which showed that the use of a flexible SG reduced the incidence of SG thrombosis.

*TRB* study also confirmed greater flexibility of A-SG, particularly for angles greater than 90°. If the SG has an excessive bending rigidity (or a high *TRB*) and is deployed in a tortuous arterial system, endoleaks can appear since the device may not fit properly to the curvature of the artery (affixing issue). In our study, Z-SG *TRB* was the highest especially for large $\alpha$.

However, no equivalent clinical value was found for *TRB*. Thus, it cannot be inferred that the bending rigidity of a device is too large and that this device cannot be used for a given $\alpha$.

Stresses in stents were twice as much with Z-SG compared to A-SG, for important angulations. Z-SG may be more prone to stent fractures, although this hypothesis needs to be confirmed by clinical data.

Strains in fabric were also higher with Z-SG, particularly at the inner curvature of the SG for maximal values of $\alpha$. Fabric tear may be more frequent with Z-SG. Although very rare in clinical report, such complication with Z-SG has already been reported (Wanhainen et al., 2008).

The present study is the first report of numerical simulation of currently marketed SGs. It allowed qualitative as well as quantitative assessment of stents and graft mechanical behavior with unprecedented precision level provided to take into account the anisotropic behavior (see section 2.2.2.2) and the bending rigidity of the textile.

This methodology may set new standards in the field of SG preclinical evaluation. It could be added to requirements for SG premarket evaluation, in order to assess performance in angulated settings.

This methodology could also form the basis for SG optimization program, by providing performance assessment of enhanced designs even before the construction of prototypes.

Several limitations of this study may be mentioned. Displacement along $z$-axis was let free, and this did not correspond to the in vivo setting where extremities of the SG are deployed at fixed point of the vasculature. Therefore, $\alpha$ should be viewed as an indicator of SG deformation rather than the angulation of the iliac artery itself. Sutures between stents and graft were not modeled because this would have increased dramatically the complexity and duration of calculations. Moreover, the accuracy of the numerical SG models should be validated using experimental studies.

## Conclusion

This study confirmed that stent design strongly influences bending behavior of aortic stent-grafts by demonstrating greater flexibility of spiral stented over Z-stented devices. The use of flexible stent-grafts may decrease the incidence of complications in the setting of tortuous aorto-iliac aneurysms. Numerical simulations could be used to compare flexibility of other commercially available stent-grafts as well as testing newly designed devices.

# TABLES

**Table 1**     Geometrical and computational features of stent-grafts

|  | Aorfix A-SG | Zenith Z-SG |
|---|---|---|
| *Degrees Of Freedom (DOF)* | 86604 | 75564 |
| **Graft** | | |
| Diameter (mm) | 16 | 16 |
| Length (mm) | 88 | 82 |
| Number of elements | 24696 | 23016 |
| **Stents** | | |
| Stent height (mm) | 88 | 12 |
| Number of stents | 1 | 5 |
| Wire radius (mm) | 0.125 | 0.14 |
| Number of elements | 2000 | 1000 |

**Table 2**     Material properties of Nitinol

| Parameters | Values |
|---|---|
| Austenite elasticity $E_A$ (MPa) | 40000 |
| Austenite Poisson's ratio $v_A$ | 0.46 |
| Martensite elasticity $E_M$ (MPa) | 18554 |
| Martensite Poisson's ratio $v_M$ | 0.46 |
| Transformation strain $\varepsilon^L$ | 0.04 |
| Start of transformation loading $\sigma_L^S$ (MPa) | 390 |
| End of transformation loading $\sigma_L^E$ (MPa) | 425 |
| Ultimate tensile strength $\sigma_R$ (MPa) | 827 – 1172 |

**Table 3**     Material properties of 316L stainless steel

| Parameters | Values |
|---|---|
| Young's modulus $E$ (MPa) | 196000 |
| Poisson's ratio $v$ | 0.3 |
| Yield stress $\sigma_S$ (MPa) | 205 |
| Ultimate tensile strength $\sigma_R$ (MPa) | 490 - 690 |
| Ultimate strain $\varepsilon_R$ | 0.6 |

**Table 4**     Material properties of PET

| Parameters | Values |
|---|---|
| $E_{\theta=0°} = E_L$ (MPa) | 225 ± 10% |
| $E_{\theta=90°} = E_C$ (MPa) | 1000 ± 10% |
| $v_{LC}$ | 0.2 |
| $G$ (MPa) | 3.6 |
| Longitudinal ultimate strain $\varepsilon_R^L$ | 0.23 |
| Circumferential ultimate strain $\varepsilon_R^C$ | 0.18 |

**Table 5**     Stent spacing variation for both SGs on inner and outer curvatures

| | A-SG | | Z-SG | |
|---|---|---|---|---|
| $\alpha$ | Inner curvature: AB (mm) | Outer curvature: A'B' (mm) | Inner curvature: CD (mm) | Outer curvature: C'D' (mm) |
| *0°* | 4.4 | 4.4 | 5.5 | 5.5 |
| *90°* | 4.2 | 4.4 | 0.0 | 5.5 |
| *180°* | 1.7 | 4.4 | 1.0 | 4.5 |

**Table 6**     Cross-section shape change in *x'y'*-plane for both SGs

| | A-SG | | | Z-SG | | |
|---|---|---|---|---|---|---|
| $\alpha$ | $d_{x'}$ (mm) | $d_{y'}$ (mm) | $d_{x'}/d_{y'}$ | $d_{x'}$ (mm) | $d_{y'}$ (mm) | $d_{x'}/d_{y'}$ |
| *0°* | 16.0 | 16.0 | 1.0 | 16.0 | 16.0 | 1.0 |
| *90°* | 16.0 | 16.0 | 1.0 | 16.9 | 14.2 | 1.2 |
| *180°* | 16.6 | 15.5 | 1.1 | 19.8 | 6.4 | 3.1 |

# FIGURES

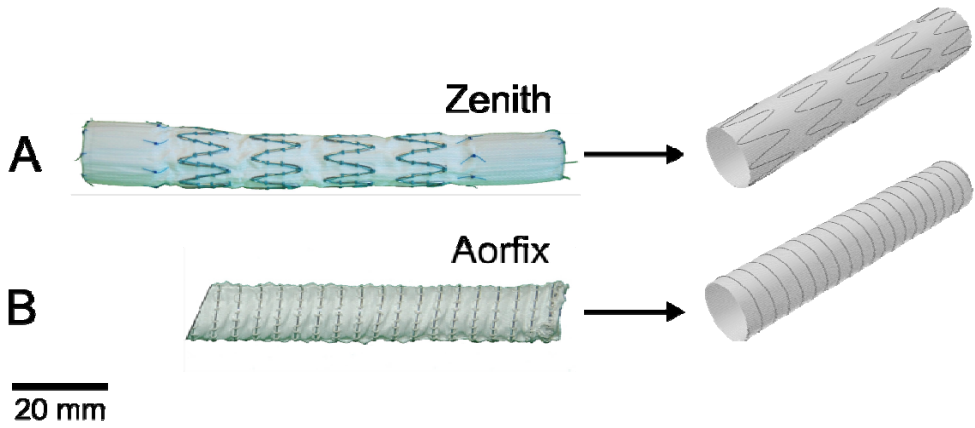

**Figure 1:** Selected manufactured stent-graft limbs and their numerical models. A: Zenith *(Cook Medical Europe)*, B: Aorfix *(Lombard Medical)*.

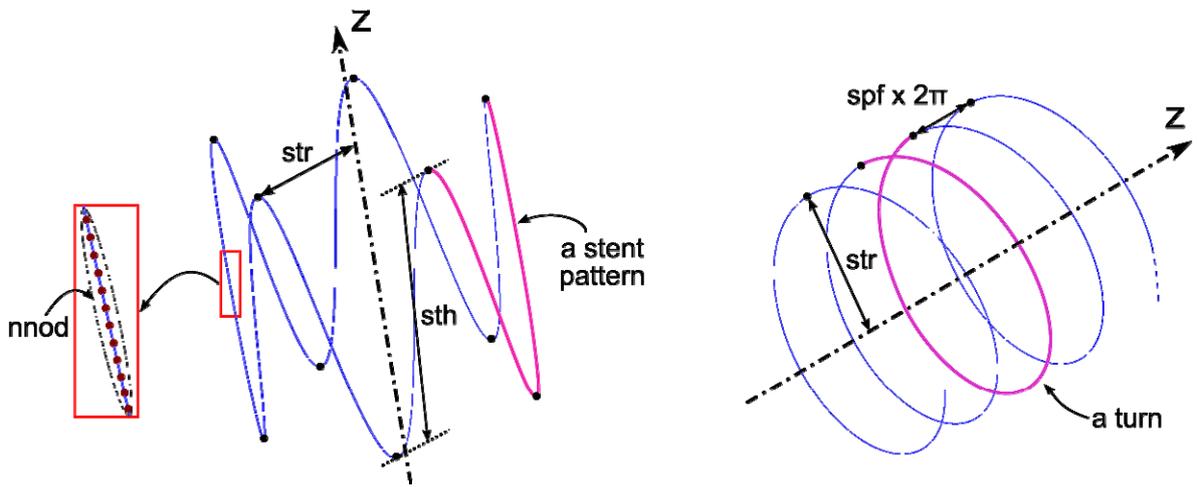

**Figure 2:** Parameters used for stent modeling and meshing.

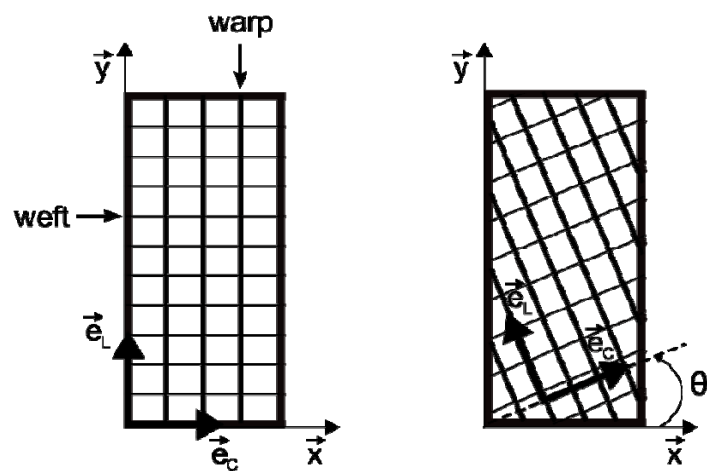

**Figure 3:** Orientation of fabric samples (*θ*: orientation angle)

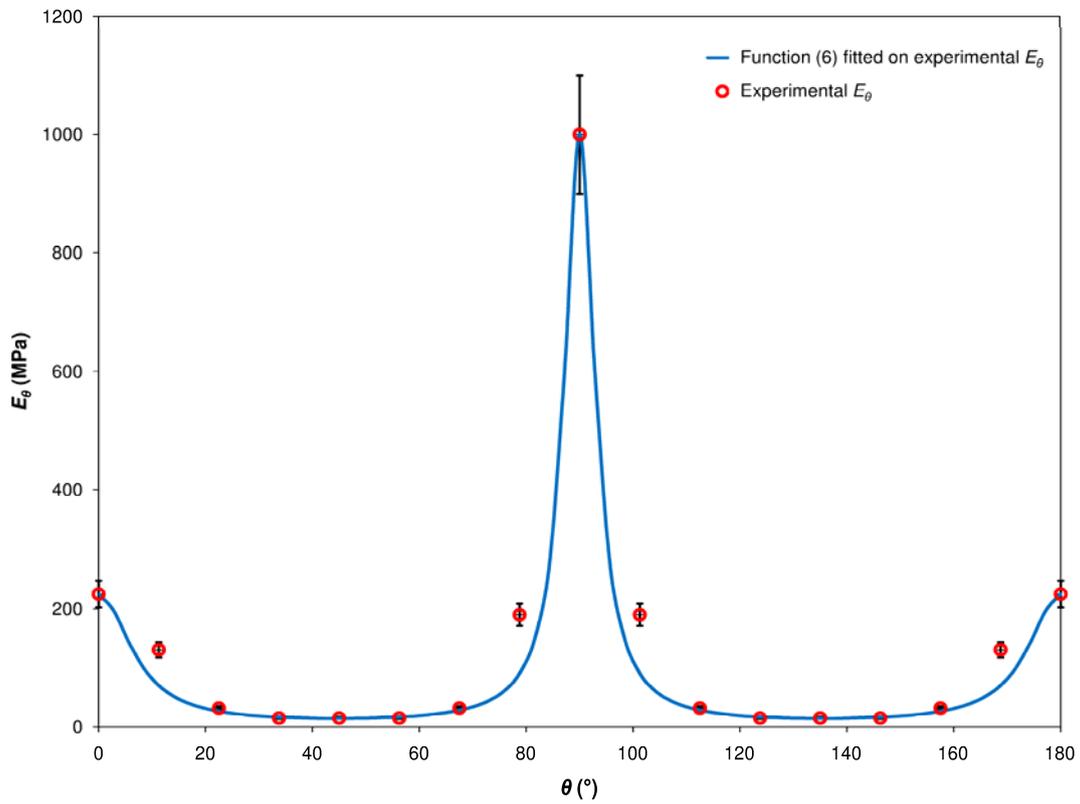

**Figure 4:** Variations of the textile elastic modulus $E_\theta$ with respect to angle $\theta$ defined as the angle between the tensile direction and the yarn direction.

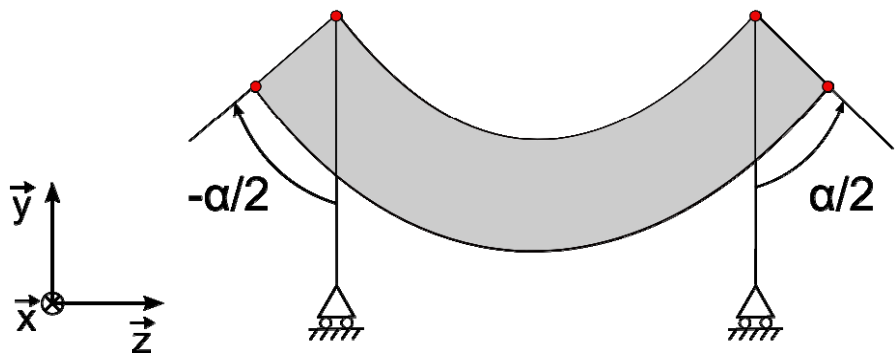

**Figure 5:** Schematic view of the boundary conditions used for stent-graft bending ($\alpha$: bending angle).

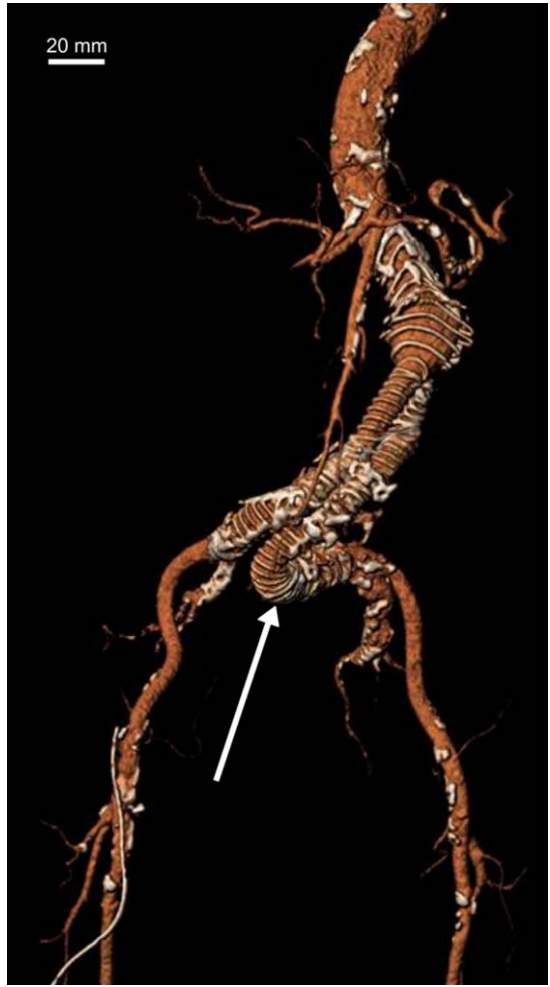

**Figure 6:** Patient with angulation of about 180° of the left common iliac artery.

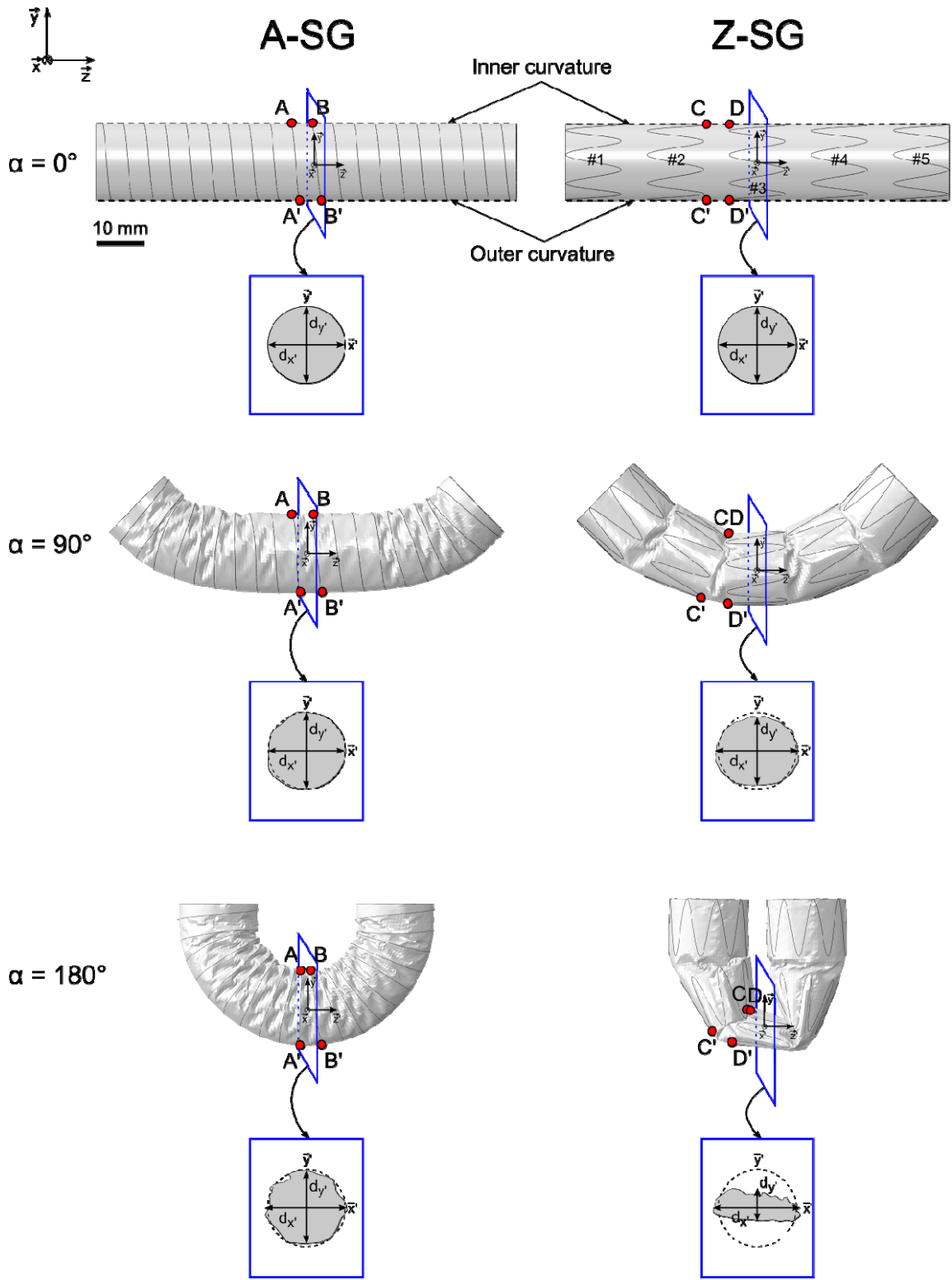

**Figure 7:** Assessment of global bending behavior of stent-grafts for $\alpha$ ranging from 0° to 180°. Plots A, B, A' and B' (C, D, C' and D', respectively) allowed to assess stent spacing variation for A-SG (Z-SG, respectively) on inner and outer curvatures. Distances $d_{x'}$ and $d_{y'}$ allowed to assess the shape change of the cross-section in $x'y'$-plane.

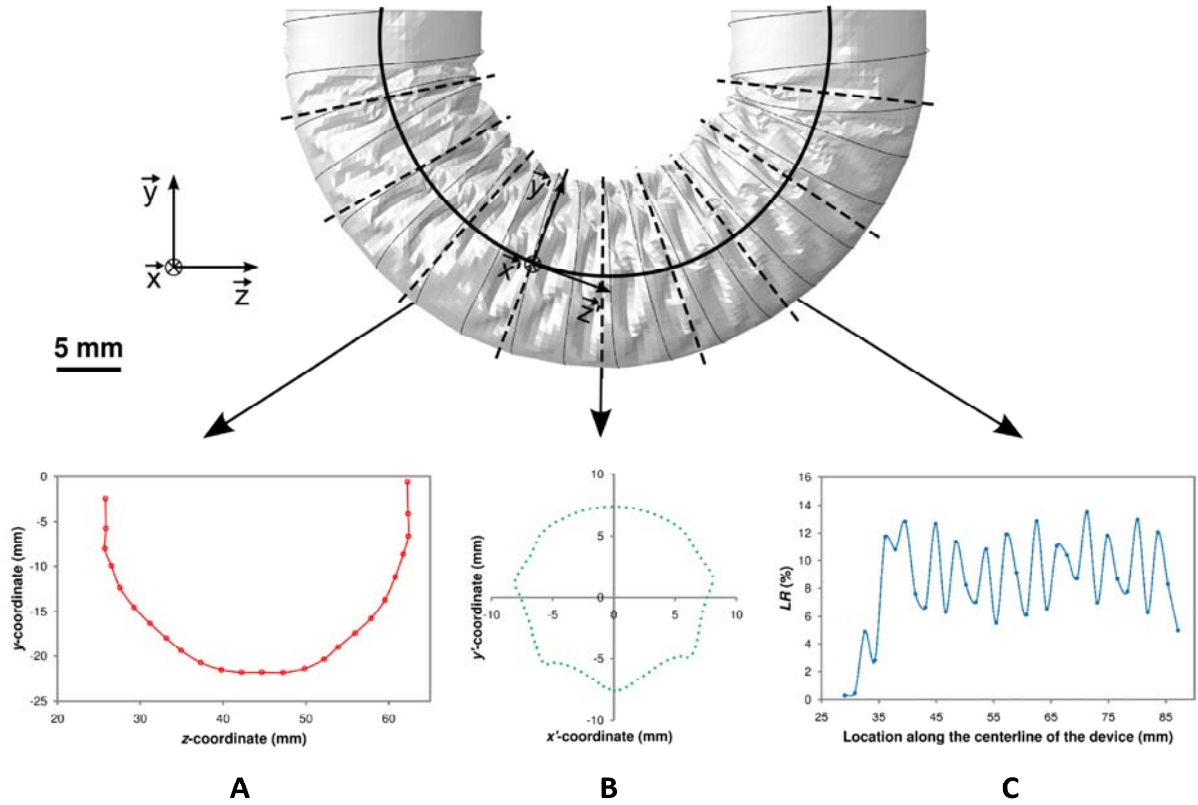

**Figure 8:** Computation of stent-graft cross-sectional area. A: Location of the centroïds forming the centerline of the stent-graft. B: Projection of the cross-section in the $x'y'$-plane. C: *LR* as a function of the location along the centerline.

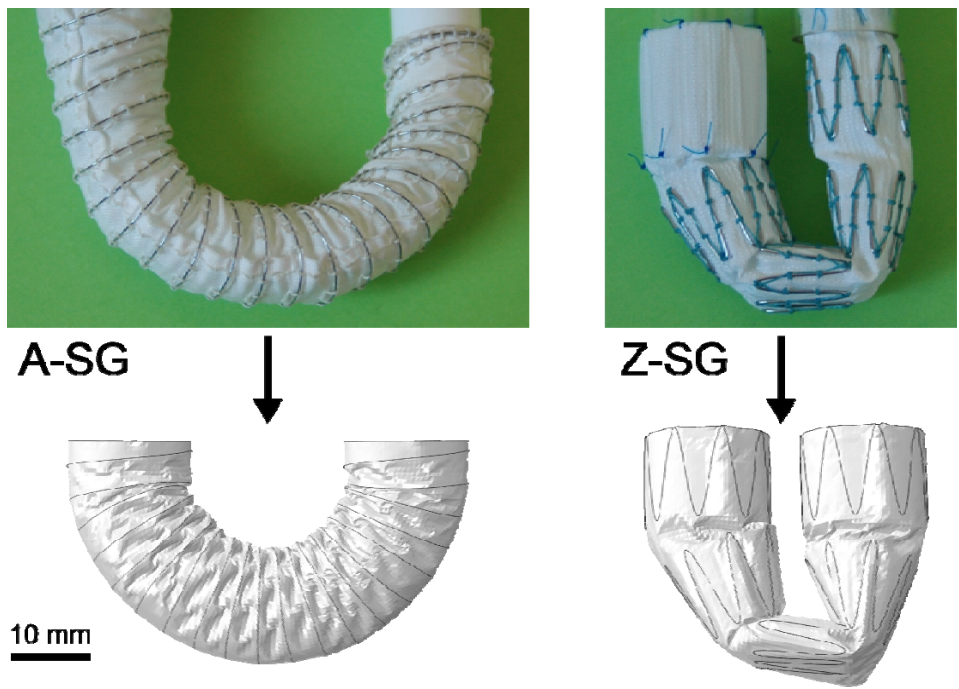

**Figure 9:** Consistency of global deformation at 180° between actual samples and numerical models.

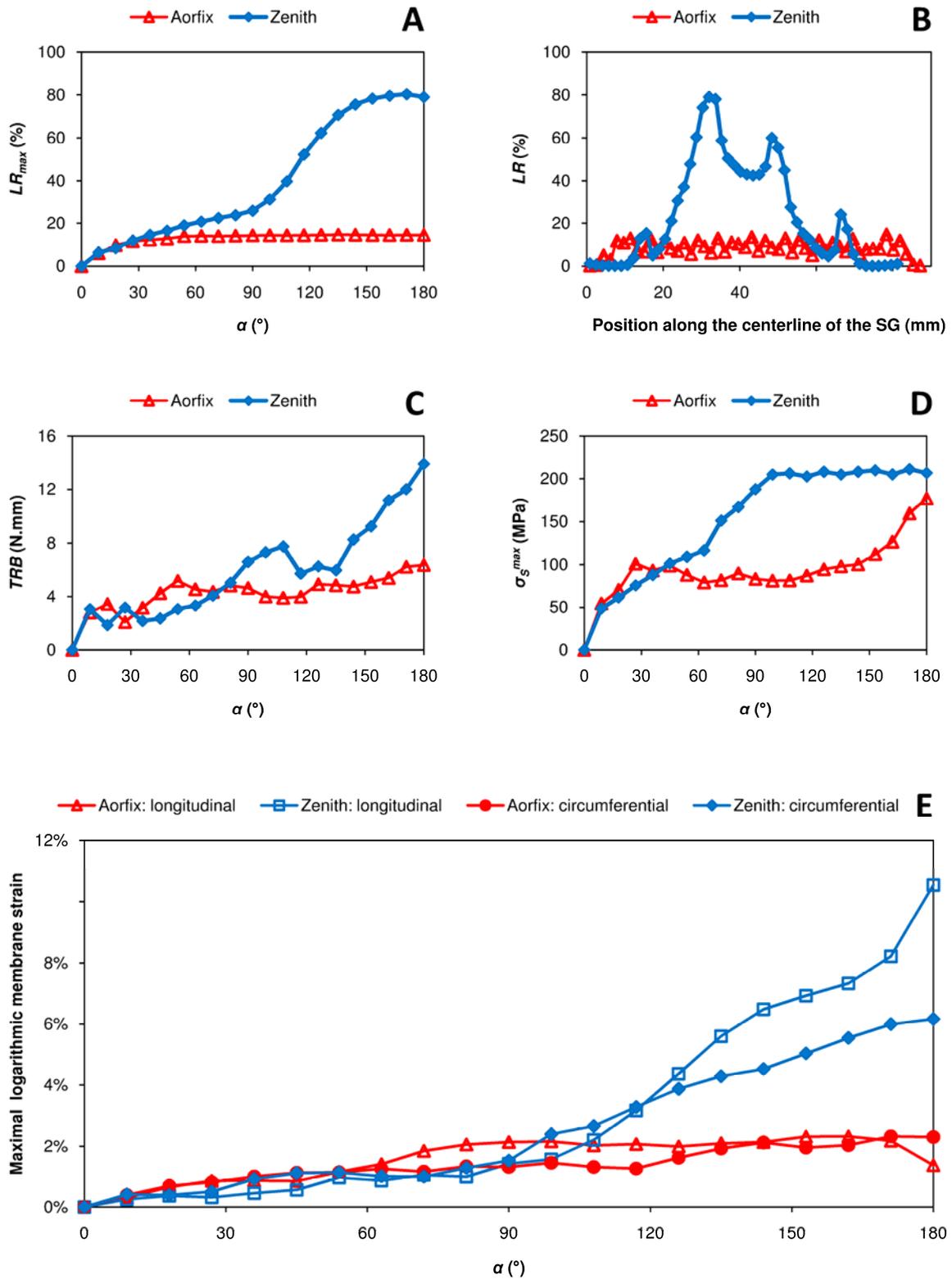

**Figure 10:** Quantitative assessment of SGs flexibility, stresses in stents, and strains in fabric. A: $LR_{max}$ vs. $\alpha$. B: $LR$ vs. location of the cross-section along the centerline of the stent-graft. C: $TRB$ vs. $\alpha$. D: $\sigma_S^{max}$ vs. $\alpha$. E: $\varepsilon_{LG}^{max}$ and $\varepsilon_{CG}^{max}$ vs. $\alpha$.

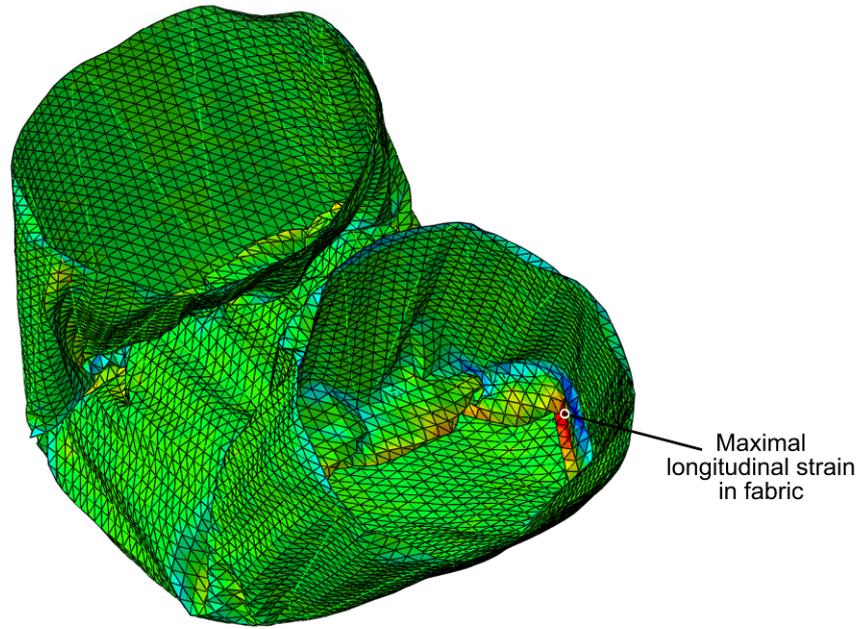

**Figure 11:** Inner view of Z-SG at 180° showing areas of high strain on fabric between stents at the level of internal curvature.